\documentclass[reprint,secnumarabic,amssymb, nobibnotes, aps, pre,groupedaddress,superscriptaddress]{revtex4-1}

\usepackage{graphicx}
\usepackage{dcolumn}
\usepackage{bm}
\usepackage{marginnote}
\usepackage{url}
\usepackage{color}

\usepackage{setspace}

\begin{document}

\preprint{xxxx}

\title{Slowdown of interpenetration of two counterpropagating plasma slab due to collective effects}

\author{N. Shukla}
\email{nshukla@tecnico.ulisboa.pt}
\affiliation{CINECA High-Performance Computing Department, Casalecchio di Reno, Italy}
\affiliation{GoLP/Instituto de Plasmas e Fus\~ao Nuclear, Instituto Superior T\'ecnico, Lisbon, Portugal}%
\author{K. Schoeffler}
\email{kevin.schoeffler@tecnico.ulisboa.pt}
\affiliation{GoLP/Instituto de Plasmas e Fus\~ao Nuclear, Instituto Superior T\'ecnico, Lisbon, Portugal}%
\author{J. Vieira}
\affiliation{GoLP/Instituto de Plasmas e Fus\~ao Nuclear, Instituto Superior T\'ecnico, Lisbon, Portugal}%
\author{R. Fonseca}
\affiliation{GoLP/Instituto de Plasmas e Fus\~ao Nuclear, Instituto Superior T\'ecnico, Lisbon, Portugal}%
\affiliation{DCTI/ISCTE, Instituto Universitario de Lisboa, Lisbon, Portugal}%
\author{E. Boella}
\affiliation{Department of Physics, University of Lancaster, Lancaster, UK}%
\affiliation{The Cockcroft Institute, Sci-Tech Daresbury, Warrington, UK}%
\author{L. O. Silva}
\email{luis.silva@tecnico.ulisboa.pt}
\affiliation{GoLP/Instituto de Plasmas e Fus\~ao Nuclear, Instituto Superior T\'ecnico, Lisbon, Portugal}%

\date{\today}

\begin{abstract}
    The nonlinear evolution of electromagnetic instabilities driven by the interpenetration of two $e^-\,e^+$ plasma clouds is explored using {\it ab initio} kinetic plasma simulations. We show that the plasma clouds slow down due to both oblique and Weibel generated electromagnetic fields, which deflect the particle trajectories, transferring bulk forward momentum into transverse momentum and thermal velocity spread. This process causes the flow velocity $v_{inst}$ to decrease approximately by a factor of $\sqrt{1/3}$ in a time interval $\Delta t_{\alpha B} \omega_p \sim c/(v_{fl}\sqrt{\alpha_B})$, where $\alpha_B$ is the magnetic equipartition parameter determined by the non-linear saturation of the instabilities, $v_{fl}$ is the initial flow speed, and $\omega_p$ is the plasma frequency. For the $\alpha_B$ measured in our simulations, $\Delta t_{\alpha B}$ is close to $10 \times$ the instability growth time. We show that as long as the plasma slab length $L > v_{fl} \Delta t_{\alpha B}$, the plasma flow is expected to slow down by a factor close to $1/\sqrt{3}$.
\end{abstract}

\pacs{xxxxxx}
\maketitle

\section{Introduction}
The interaction of pair-plasmas is inevitably present in many astrophysical conditions such as supernova explosions, gamma ray bursts, TeV-Blazars, etc \cite{Weinberg72, Gibbons83, Piran04}. One of the unanswered mysteries of the early universe has been to identify a mechanism that explains the origin and evolution of magnetic fields starting from essentially zero \cite{Schlickeiser03, Uzdensky14, Widrow, Kulsrud-2008}. The Weibel instability is a mechanism that is well accepted to play a role, and has been intensively studied both theoretically as well as in laboratory plasma in collisionless regimes \cite{Medvedev-ApJ-1999, Schoeffler-2014, Shukla20, Silva20, Arrowsmith, Peterson21}. The corresponding growth rates range from a few microseconds to a few tenths of a second, consistent with the time scales of GRBs. These instabilities arise due to an anisotropic velocity distribution in the plasma (WI) or due to a counter-streaming flow of plasma slabs (CFI) \cite{Weibel, Fried, Tsilva}.  The impact of beam-plasma instabilities upon gamma-ray emission in bright TeV sources and their subsequent cosmological consequences have been previously investigated theoretically and numerically using realistic parameters.

Several experiments have been performed where a pair plasma is generated in the laboratory to study the spatiotemporal evolution of Weibel generated magnetic fields \cite{Stamper-1971, Nilson, Sari-Nature-2015, Gode-2017}. However, plasma instabilities are well understood analytically in the linearized regime with small perturbations to an infinite homogeneous plasma. Furthermore, the theoretical estimates of the linear growth of plasma instabilities, which generate
electromagnetic fields during the interaction of two plasma slabs, complemented by numerical simulations, have been well studied~\cite{Luis-2002,Shukla-JPP-2012, Ruyer16}. The electromagnetic fields driven by these plasma instabilities will lead to the bulk slowdown of the counterpropagating plasma slabs as long as there is sufficient time for the instabilities to grow. The exact time required before a significant slowdown occurs, however, depends on nonlinear effects not included in such analysis and can be only captured via numerical simulations, as performed in this paper. Our studies, therefore, can place a limit on the interaction strength by considering the full nonlinear dynamics associated with these instabilities.

In this work, we consider the simple possibility of an electromagnetic self-interaction between two collisionless $e^-, e^+$ plasma slabs. Collisionless plasma dynamics is both a well-studied field and an area of active research with rich dynamics that are still not fully understood. It is known that two counterpropagating plasmas are subjected to several microinstabilities that generate growth of electromagnetic fields, involving transverse and parallel modes. The full unstable wavenumber k spectrum has been intensively studied in the cold plasma limit \cite{Bret10, Bret2006}. There are three main dominating instabilities which exist with different wave-vectors with respect to the flow. First, the two-stream instability (TSI), which has a wave-vector aligned with the flow, is driven by the two peaked nature of the velocity distribution \cite{Bohm49, Fainberg70}. Second, anisotropy in the velocity spread in different directions (larger along the flow direction) excites the Weibel/Filamentation instability (WI/FI) with a wave-vector normal to the flow \cite{Weibel, Sari-Nature-2015, shukla18}. Finally, a hybrid of these two modes, with a wave vector with and angle oblique to the flow, is known as the oblique instability (OBI) \cite{Bret2006}.

The dimensionless quantities, $\alpha_B = U_B/\mathcal{E}_p$ and $\alpha_E = U_E/\mathcal{E}_p$ are the respective magnetic and electric equipartition parameters. Here the energy in the magnetic fields  ($U_B = \int B^2 dV$) and electric fields ($U_E = \int E^2 dV$) is normalized to the initial total kinetic energy in the system $\mathcal{E}_p =\sum_\alpha \int n_{0\alpha} m_e v_{fl}^2/2 dV$, summing over each species $\alpha$; in our case the number of species is $N_{SP} =2,(e^-,e^+)$. Here $m_e$, $n_{0\alpha}$, and $v_{fl}$, are the respective mass, density, and velocity of the species, and $V$ is the total volume of the two slabs. Above quantities will be used demonstrating the slowdown process that occurs during the interaction of two plasma slabs. 

\begin{figure}[t]
	\begin{center}
		\includegraphics[width=0.45\textwidth]{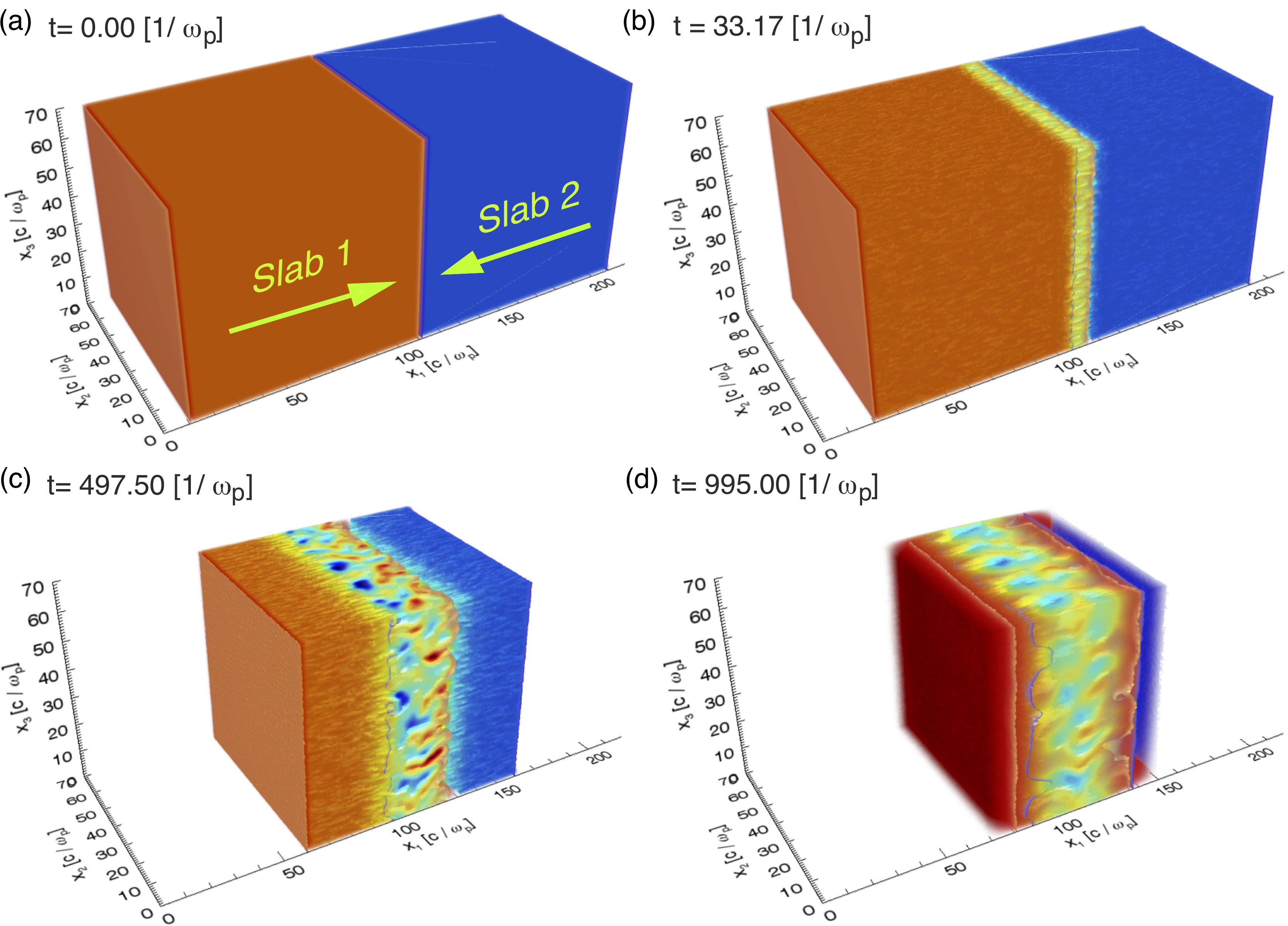}
		\caption{(color online) The temporal evolution of the $e^{-}$ filament density for simulation $R_1$, where Slab 1 (orange) and Slab 2 (blue) are moving to the respective right and left at $t \omega_p = 0.0, 33.17, 497.50, 995.00$ (a)-(d). The green color represents where the plasma from both slabs overlap.}			
		\label{Fig1}
	\end{center}
\end{figure}

\section{Computational Model}
In order to demonstrate the slowdown of plasma particles, we simulate the interaction of two initially unmagnetized electron-positron plasma slabs propagating towards each other. Two and three dimensional numerical simulations were performed with the fully relativistic, massively parallel, particle-in-cell (PIC) framework OSIRIS \cite{R1, R2}, which simulates kinetic plasmas {\it ab initio} (see Appendix A for an explanation of the PIC technique). The box has dimensions of $2.2 L\times L_{\perp}^2$ with a resolution $\Delta x$, where $L$ is the length of the plasma slab and $L_{\perp}$ is the transverse dimension. The two slabs, shown in Fig.\,\ref{Fig1}a, consist of a plasma with uniform density $n_0$ moving to the right (red) between $x_1= 0.1L-1.1L$, and moving to the left (blue) between $x_1= 1.1L-2.1L$ with a bulk proper fluid velocity $v_{fl}$.  We chose a step function density profile rather than Gaussian in order to maximize the interaction time where the plasmas are unstable, because a Gaussian profile would cause a delay of the instability onsets. We ran each simulation for one crossing time $\tau_{c} = L/v_{fl}$ with a temporal resolution $\Delta t\,\approx 0.98\Delta x/\sqrt{D}c$, where D is the number of dimensions of the simulation. 

We performed one 3D simulation ($R_1$) with 4 particles per cell for each plasma species, and 2D simulations ($R_2$, $R_5$, and $R_6$) with 16 particles per cell for each plasma species, all with $L = 100 [c/\omega_p]$, $L_{\perp} = 70 [c/\omega_p]$, and $\Delta x = 0.1 [c/\omega_p]$, where $c/\omega_p$ is the electron skin depth, $c$ the speed of light, $\omega_p = \sqrt{4 \pi e^2 n_0/m_e}$ the electron plasma frequency, e the elementary charge, and $m_e$ the electron mass. We performed more 2D simulations, also with 16 particles per cell for each plasma species, varying the slab length $L$. Simulations ($R_3$ and $R_7$) have $L = 5 [c/\omega_p]$, $L_{\perp} = 5 [c/\omega_p]$ with resolution $\Delta x = 0.01 [c/\omega_p]$, and ($R_4$ and $R_8$) have $L = L_{\perp} = 0.02 [c/\omega_p]$ with $\Delta x = 0.0004 [c/\omega_p]$.

We performed simulations varying the respective flow and thermal velocities $v_{fl}/c \in [0.01-0.1]$ and $v_{th}/c \in [0.001-0.1]$ which have been listed in Table \ref{1} and \ref{2}. Absorbing boundary conditions have been used for the fields and the particles in the direction parallel to the flow velocities, while the conditions are periodic in the transverse direction. To suppress numerical heating, a fourth order interpolation scheme has been used together with a 5-pass filter to evaluate current and fields. Larger transverse box sizes, higher spatial and temporal resolution and the higher number of particles per cell were tested, showing overall convergence.

\section{Interpretation of simulation results}
Here we first report the results from the fiducial three dimensional simulation ($R_1$). In Fig.\,\ref{Fig1}, we show four representative times over the period of one crossing time $\tau_{c} = L/v_{fl}$. During this time, the plasmas penetrate (see in Fig.\,\ref{Fig1}b), filament (see in \,Fig.\ref{Fig1}c), and  slow down significantly by $\tau_{c}$ (see in Fig.\,\ref{Fig1}d). The slowdown and isotropization of the velocity distribution occurs at three time scales, that of the two-stream/oblique instability, the Weibel instability,  and  the  crossing  time  of  the  plasma slabs.

\begin{figure}[tp!]
	\begin{center}
		\includegraphics[width=0.48\textwidth]{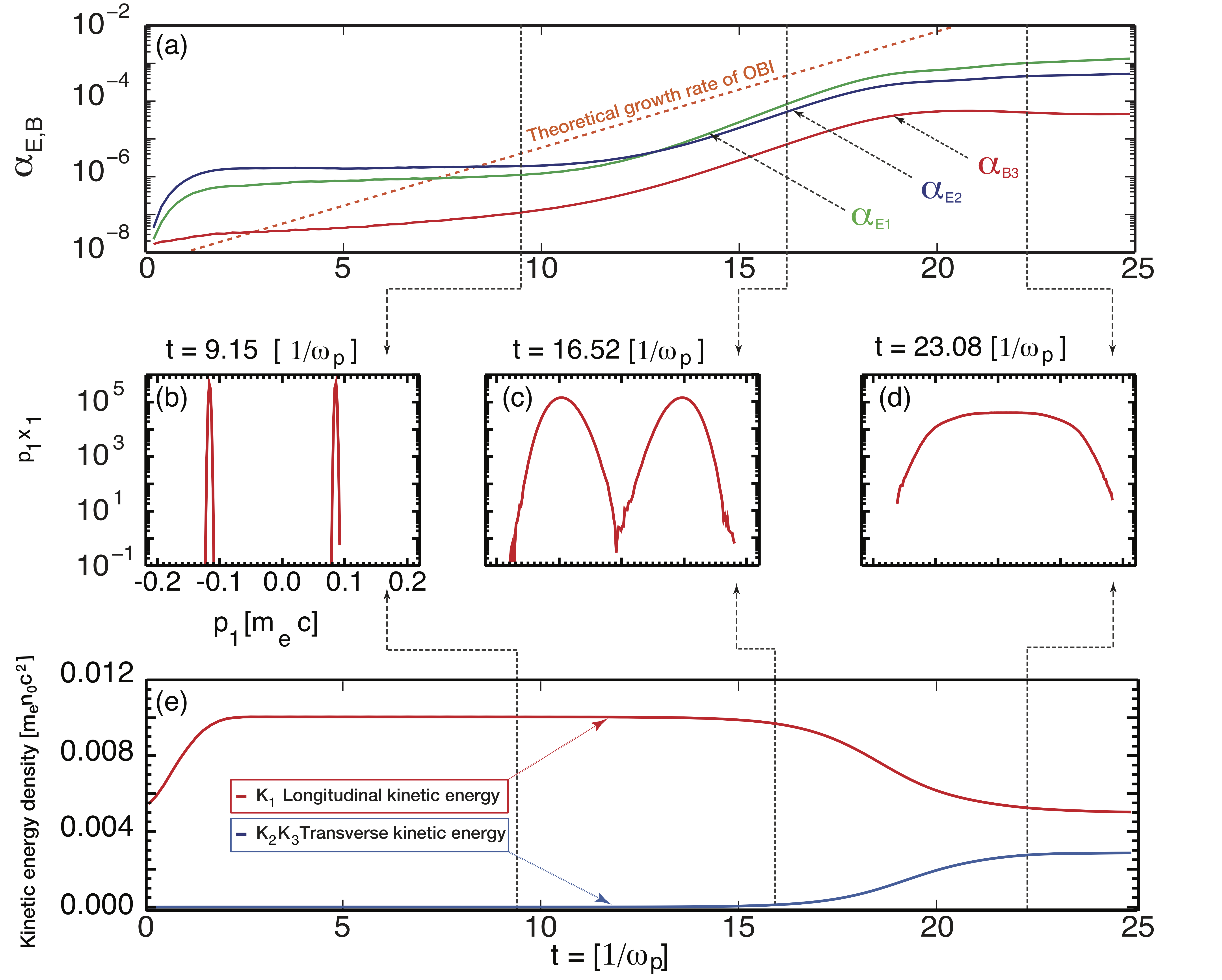}
		\caption{(a) Temporal evolution of the total electromagnetic energy of the system for the 3 components $E_{x1}$ (blue), $E_{x2}$ (green), and $B_{x3}$ (red) in their normalized form $\alpha_{E,B}$. Panels (b)-(d) show the electron distributions of momentum along the $x_1$ direction calculated between $x_1 = 109.9 - 110.1 \,[c/\omega_p]$ over all $x_2$ and $x_3$, for $t\,\omega_p = 9.15,16.52$, and $\mathrm{23.08}$. Panel (e) shows the temporal evolution of the mean kinetic energy density of electrons in slab 1 over the same region ($K_1$ (red) and  $K_2=K_3$ (blue), where  $K_1$ is the kinetic energy in the direction of flow velocity, and $K_2$ and $ K_3$ are transverse).}\label{Fig2}
	\end{center}
\end{figure}

At the earliest stage, the overlapping plasma slabs result in two peaks in velocity space in opposite directions, which drives the oblique instability. The oblique instability generates electric and magnetic fields at the expense of the initial bulk energy: a fraction of the initial kinetic energy $\epsilon_p$ is transferred into the different components of electromagnetic fields; the longitudinal electric field $E_1$ and the transverse electric and magnetic fields $E_2$, $B_3$ shown in Fig.\,\ref{Fig2}a. In Fig.\,\ref{Fig2}a, we present the temporal evolution of the electric and magnetic field energy. Throughout this study, when we refer to the oblique instability, the two-stream component of the oblique instability dominates. The oblique instability (not just two-stream) can be seen in  Fig.\,\ref{Fig2}a, since the magnitude of the longitudinal $E_1$ and the transverse electric fields $E_2$  (unique to the oblique instability) are about equal. During the linear stage of the instability at time $t \omega_p \approx 13$, the transverse electric fields $E_2$ are greater than the transverse magnetic fields $B_3$, consistent with the modes of the oblique instability. The theoretical growth rate of the oblique instability $ \Gamma_{TS} \approx \omega_p/\sqrt{2}$ \cite{Bret10,Bret2006} shown in Fig.\,\ref{Fig2}a (red dotted), matches well with the simulation result. 

To further illustrate that the instability acts like the two-stream instability, we examine the evolution of the distribution $f_0 (p_{1})$ through the course of the growth and saturation of the oblique instability in the region $\mathrm{x1 = [109.9-110.1]}\,[c/\omega_p]$. At $t\,\omega_p= 9.15$, once the two slabs overlap, the two peaks in velocity space in opposite directions are illustrated in Fig.\,\ref{Fig2}b. During the interaction, the electric field grows linearly due to the oblique instability causing a strong heating which broadens the initial particle distribution (see Fig.\,\ref{Fig2}c). After the linear phase, at time $t\,\omega_p = 23.08$, the instability saturates and the distribution is completely thermalized (see Fig.\,\ref{Fig2}d). Here the free energy that drives the oblique (two-stream) instability is no longer present. During the linear phase of the instability, the transverse electric field accelerates the particles in the transverse directions $x_2$ and $x_3$. The particles slow down in the longitudinal ($x_1$) direction, such that the total kinetic energy does not change much, as only 0.1 \% of the initial flow kinetic energy is converted into field energy, see in Fig.\,\ref{Fig2}(a,e). In this process, the kinetic energy becomes significantly isotropized (see Fig.\,\ref{Fig2}e); each of the components of the kinetic energy approach a number close to $0.004$ $[m_e n_0 c^2]$. 

Here we estimate the typical timescale for isotropization of the kinetic energy in D dimensions solely due to the electric fields. Assuming that the total kinetic energy does not change, complete isotropization of the velocity distribution will occur once the change of velocity along the directions perpendicular to the flow is

\begin{equation} \label{eq:1}
    \Delta v = \sqrt{\left< v_i^2 \right>} \approx 1/\sqrt{D} v_{fl} \text{.}
\end{equation}

Each component of the electric field can be expressed as $E_i = \omega_p m_e/e \sqrt{\alpha_{E,loc} N_{sp}/D} v_{fl}$, where $\alpha_{E,loc} \equiv E^2/m n_0 N_{sp} v_{fl}^2/2$ is the local electric equipartition parameter. From the Lorentz equation:

\begin{equation} \label{eq:2}
\frac{\Delta v}{\Delta t} = -\frac{e}{m_e} E_i 
\end{equation} 
by substituting $\Delta v/v_{fl}$ and $E_i$, we can estimate the isotropization time due to electric field:

\begin{equation} \label{eq:3}
\Delta t_{\alpha_E} \omega_p = \sqrt{\frac{1}{\alpha_{E,loc} N_{s p}}} \approx 5.85,
\end{equation}
where $\alpha_{E,loc}$ is calculated at $t \omega_p = 16.52$, when the $\alpha_E$ begins to saturate see Fig.\ref{Fig2}a; $\alpha_{E,loc}$ reaches a maximum $\alpha_{E,loc} = 1.46 \%$ at $x_1 = 109.8\,[c/\omega_p]$. This timescale is in good agreement with the simulation result (see Fig.\,\ref{Fig2}e between $t \omega_p=16.52$ and $22.37$. Note that further isotropization occurs after $t \omega_p = 25.00$.

\begin{figure}[t!]
   \begin{centering}	
		\includegraphics[width=0.45\textwidth]{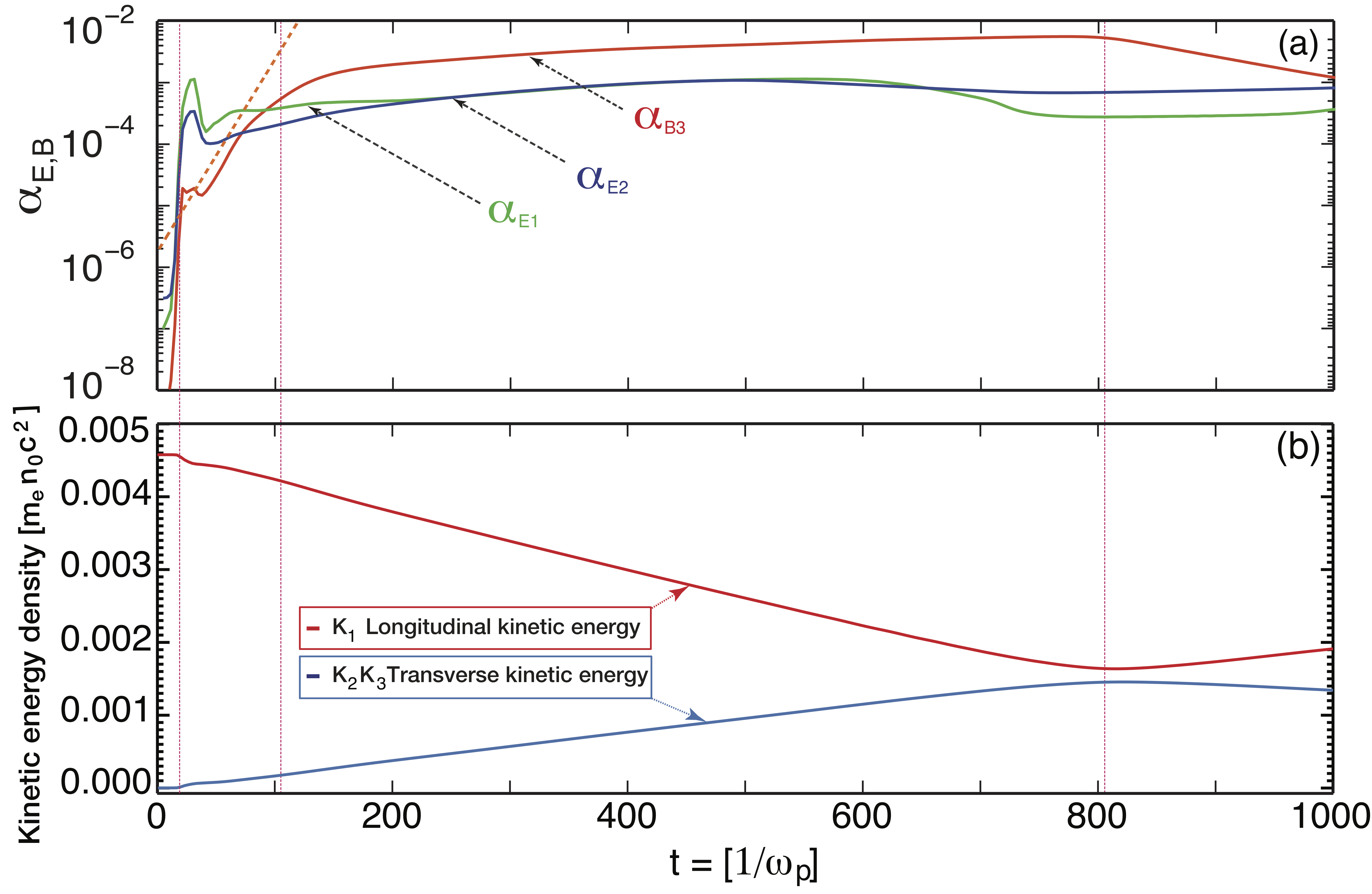}
	\end{centering}
	\caption{(a) Temporal evolution of the total electromagnetic energy of the system for the 3 components $E_{x1}$ (blue), $E_{x2}$ (green), and $B_{x3}$ (red) in their normalized form $\alpha_{E,B}$. Dashed lines indicate $\Delta t_{\epsilon E}, \Delta t_{\epsilon B}$, and the time plotted in Fig.\,\ref{Fig3}d when the maximum velocity along the plasma slab drops below 0.5$v_{fl}$ (the slabs have transited past each other). Panel (b) shows the temporal evolution of the mean kinetic energy density of electrons in slab 1 ($K_1$ (red) and  $K_2=K_3$ (blue), where $K_i\equiv \left<m_e n v_{i}^2/2 \right>/m_e n_0 c^2$, averaged over all $x_1, x_2$, and $x_3$. $K_1$ is longitudinal kinetic energy in the direction of flow velocity and, $K_2$ and $ K_3$ are transverse). The red dotted line is the theoretical growth rate of the Weibel instability.}\label{Fig4}
\end{figure}

The Weibel instability is driven by a temperature anisotropy\,\cite{Weibel}. Although the oblique instability thermalizes, and significantly isotropizes the plasma velocity distribution, an anisotropy remains (see Fig.\,\ref{Fig2}e). At about $t \omega_{p} = 40.00$, the magnetic field energy grows at a rate consistent with the theoretical growth rate ($ \Gamma_{WI}/\omega_{p} \approx ~ v_{fl}/\sqrt{2} \approx 0.07$ \cite{Bret10, Bret2006}) indicated by the red dotted in Fig. \ref{Fig4}a. The instability saturates at about $t \omega_{p} = 100.00$, and the magnetic fields are  responsible for the further isotropization of the slabs. After saturation, the magnetic field strength grows linearly, between $t \omega_{p} = 200.00- 800.00$, as the shock front propagates across the plasma slab. 
The kinetic energy in each direction is defined as $K_i\equiv \left<m_e n v_{i}^2/2 \right>/m_e n_0 c^2$. $K_1$ is the longitudinal kinetic energy in the direction of flow velocity, and $K_2$ and $ K_3$ are transverse
The magnetic field isotropizes the kinetic energy by bending the trajectories such that $K_1$ is converted to $K_2$ and $K_3$ see Fig. \ref{Fig4}b.

\begin{figure}[tp!]
	\begin{centering}	
		\includegraphics[width=0.44\textwidth]{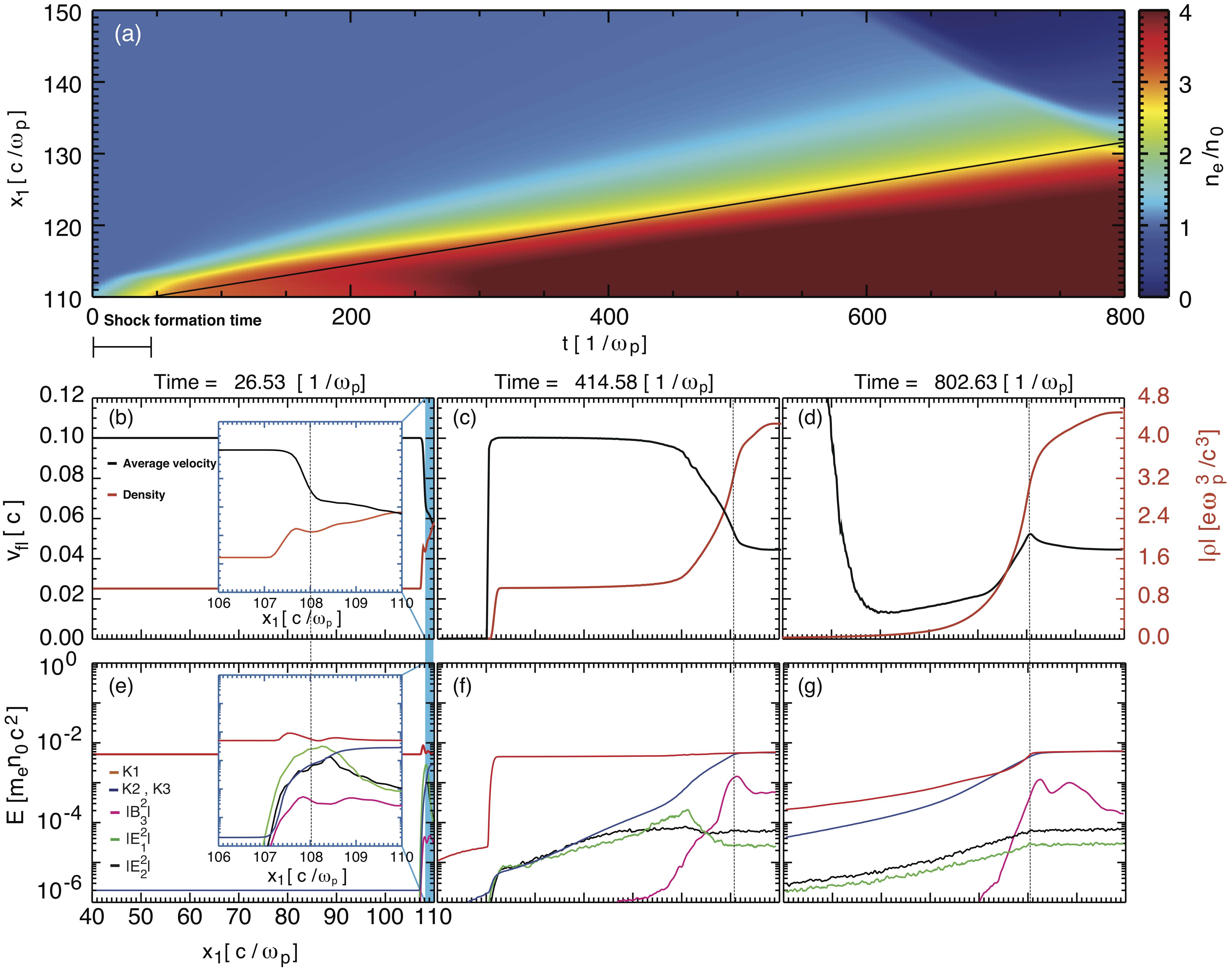}
	\end{centering}
	\caption{(color online) Panel (a) shows the evolution of the plasma density vs. time. The solid line indicates a line of best fit for the shock front ($x_1 = v_s(t-t_0)$), where $t_0 \omega_p$ is $110$ and $v_s/c =0.0286$. Panels (b-d) show the spatial profiles of the average $x_1$ velocity of electrons with positive velocity (to the right) (black) and average density (red) averaged along the $x_2$ and $x_3$ directions. Panels (e-g) show the energy densities of each component of the electromagnetic fields and the kinetic energy averaged along the $x_2$ and $x_3$ directions. The black dotted line is the position of the shock front, and at $t \omega_p = 26.53$ where there is no shock, this represents the density front.}\label{Fig3}
\end{figure}

A similar method is adopted to calculate the istropization time due to the magnetic fields ($\alpha_{B,loc} = 0.3 \% $ calculated from Fig.\,\ref{Fig3}f at $x_1 = 109.8\,[c/\omega_p]$)
\begin{equation} \label{eq:4}
\Delta t_{\alpha_B} \omega_p = \sqrt{\frac{2}{D \alpha_{B,loc} N_{s p}}} \frac{c}{v_{fl}} \approx 105.4,
\end{equation}
consistent with the saturation of the Weibel magnetic field at time
$t \omega_p \approx 100.0$.

Fig.\,\ref{Fig3}a illustrates the formation and propagation of the shock front.
At the shock front, the density of plasma slabs increases from $n_0$ to 4$n_0$ (see also Fig.\,\ref{Fig3}b-d). 
The solid line indicates a line of best fit of the shock front $x_1 = v_s(t-t_0)$, where the shock formation time $t_0 \omega_p$ is $110$ and the shock propagation velocity is $v_s/c = 0.0286$. The shock develops fully once both the oblique (two-stream) and the Weibel instabilities have time to grow and saturate. The isotropization time from Eq.\,\ref{eq:4}, $\Delta t_{\epsilon_B} \omega_p\approx 105.4$, which we have shown to be consistent with the Weibel saturation time, is also consistent with the shock formation time $t_0 \omega_p = 110$ shown in Fig.\,\ref{Fig3}a. 

The growth and saturation of both instabilities can be seen as a function of space, during various times, eventually leading to the total slowdown of the plasma slab. At early times (see Fig.\,\ref{Fig3}e) the E-fields from the oblique instability begin to isotropize the kinetic energy and significant slowdown occurs (see Fig.\,\ref{Fig3}b). The electric field of the oblique modes are coincident with the density front at $t\,\omega_p = 26.53$ (see Fig.\,\ref{Fig3}b,e). At this stage the shock is not yet formed.

As the plasma slabs overlap further, the Weibel instability begins to play a role (see Fig.\,\ref{Fig3}f). At $t\,\omega_p = 414.58$, the density front has become a fully formed shock. The slabs slow down in the shock region, as seen in Fig. \ref{Fig3}(c,f) between $L = 80-100\,[\mathrm{c/\omega_{p}}]$. This slowdown of $v_{fl}$ across the shock is stronger than in Fig. \ref{Fig3}b.  In Fig. \,\ref{Fig3}f, at the position $x_1 = 92.0 [c/\omega_p]$, the electric field generated by the oblique instability contributes to the isotropization of the kinetic energy. The magnetic field generated by the Weibel instability,  peaking at $x_2 = 102.0 [c/\omega_p]$, further isotropizes the kinetic energy (until $K_1 \approx K_2 \approx K_3$). Assuming a steady state propagating shock front has been established (evidenced in \ref{Fig3}a), and the flow velocity is $v_{fl}=0.1c$, the time corresponding to the region where slowdown occurs $x_1 = [90 -100] [c/\omega_p]$ (see \ref{Fig3}c,f), can be estimated by $\Delta x_1 /v_{fl} \approx 100$ [$1/\omega_p$], matching the Weibel saturation time. 

At the transit time $t \omega_p = 802.63$, once the shock has traversed the entire slab (see Fig. \ref{Fig3}d), the slowdown of the two slabs is completed. Fig. \ref{Fig3}g shows the complete isotropization across the entire slab. Similar studies have been performed explaining deflection and isotropization of the particles coming from upstream in the collisionless shock \cite{silva_2006, Kato_2008}. 

\begin{figure}[htp!]
		\begin{centering}	
		\includegraphics[width=0.45\textwidth]{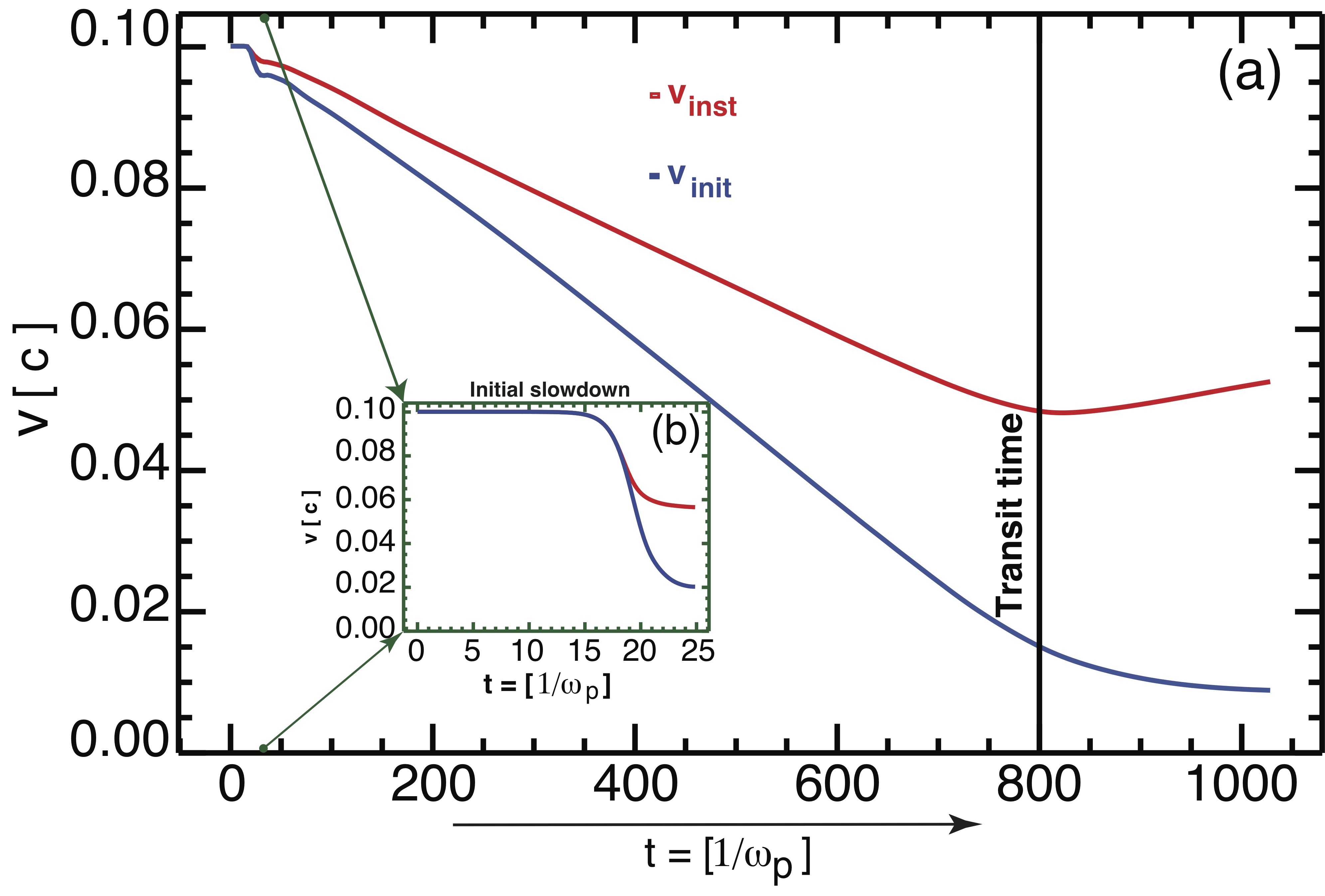}
	\end{centering}
		\caption{(a) (Color online) Temporal evolution of $v_{init}$, the average velocity over the entire simulation box of all electrons that were initially moving to the right (blue), and $v_{inst}$, the average velocity of all electrons moving to the right at time $t$ (red) and (b) averaged between $x_1 = 110.3-110.5 \,[c/\omega_p]$ over all $x_2$. The solid line is the time plotted in Fig.\,\ref{Fig3}d when the maximum velocity along the plasma slab drops below 0.5$v_{fl}$ (after the slabs have transited past each other).}\label{Fig5}
\end{figure}

We determined two measures that can be used to quantify the slowdown of the plasma; i) the average velocity of the  electrons initially moving to the right $v_{init} = \int_{-p}^{+p} p f(p) dp /\int_{-\infty}^{+\infty} f(p)dp$ (from the left slab only) (blue) and ii) the average velocity of the electrons moving to the right at a particular instant $v_{inst} = \int_{0}^{+\infty} p f(p) dp /\int_{0}^{+\infty}f(p) dp$ (from both slabs) (red). We define a significant slowdown as $v_{init},v_{inst} <0.9\,v_{fl}$ to determine whether there is slowdown or not. At the transit time $t \omega_p = 802.63$, the velocities reach $v_{init} = 0.1444\,v_{fl}$ and $v_{inst} = 0.5\,v_{fl}$, a significant slowdown under both measures (see Fig.\,\ref{Fig5}a).

Note the slowdown of $v_{inst}$ is consistent with the prediction that once isotropized $v_{inst} = \Delta v \approx 1/\sqrt{3} v_{fl} \approx 0.577 v_{fl}$ in 3D, see Eq.\,\ref{eq:1}.
Furthermore, a significant slowdown of the front of the plasma slabs (at the center of the box) is shown in the inset Fig.\,\ref{Fig5}b. This occurs at a much shorter time scale $t \omega_p = 25$, the time scale of the two-stream/oblique instability.

\begin{figure}[htp!]
    \begin{centering}	
		\includegraphics[width=0.45\textwidth]{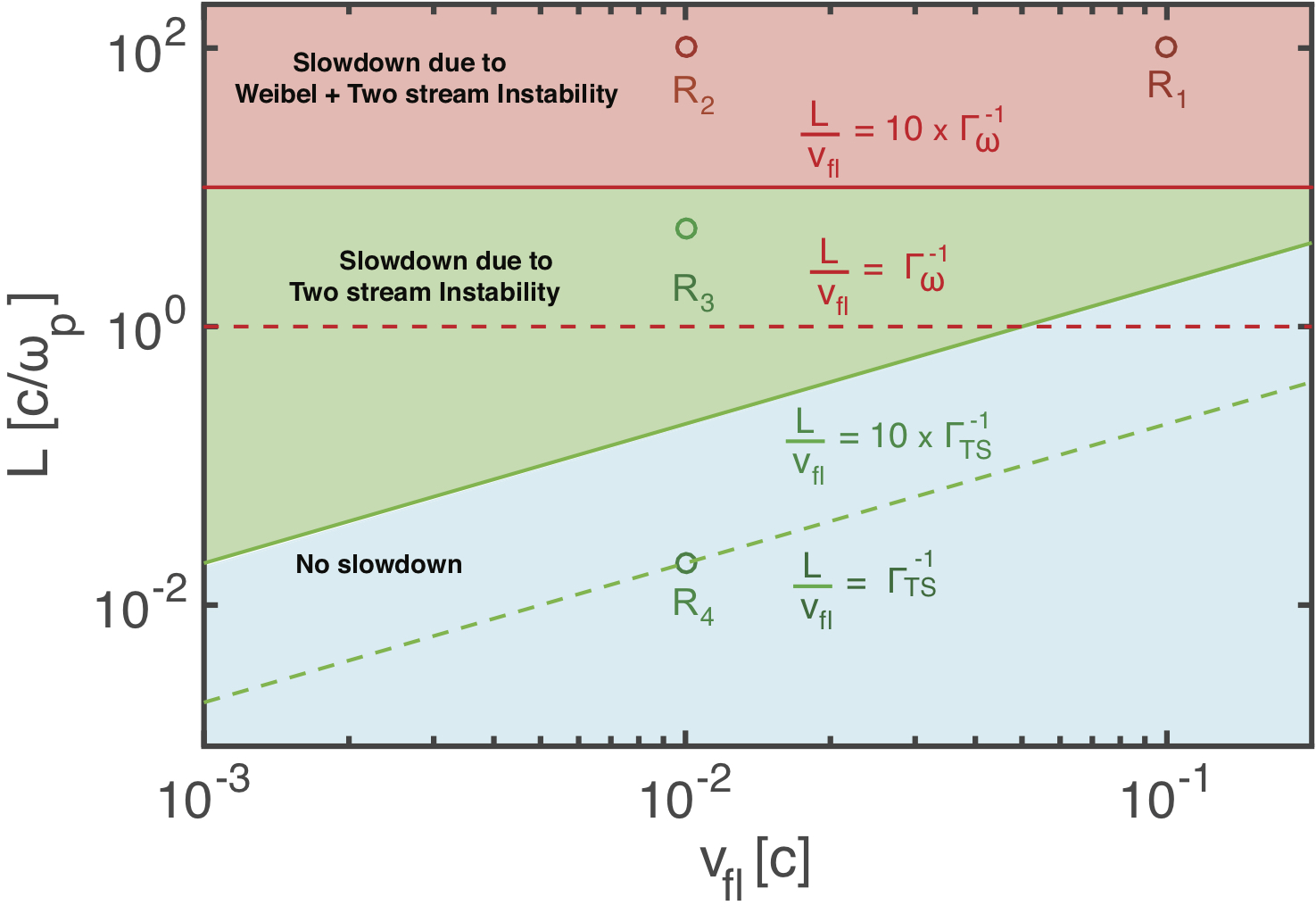}
	\end{centering}
		\caption{\emph{Length of plasma slab vs initial velocity.} Parameter space showing when the slowdown of the plasma slab is expected due to two-stream/oblique or by the Weibel instability. In the red region, both the Weibel and the two-stream/oblique instabilities are expected to lead to a significant slowdown. In the green region, only the two-stream/oblique instability should act to slow down the plasma. In the blue region slowdown is not expected. The dashed red and green lines show where the growth rate of the respective Weibel and two-stream/oblique instabilities are equal to the rate at which the two slabs cross past each other. Significant slowdown is expected after a factor of $10$ growth times, based on our predicted slowdown times $\Delta t_{\alpha B} \Gamma_{W} \approx 6.324$ and $\Delta t_{\alpha E} \Gamma_{TS} \approx 8.66$.}\label{Fig7}
\end{figure}

The relative growth of the WI/OBI and the crossing time $\tau_c$, which determines whether there can be a slowdown of the two plasma-slabs, depends on $L$, $v_{fl}$, and $v_{th}$. Now, we present results from simulations where we vary these key parameters. As there is little difference between the 2D and 3D runs, in order to minimize the computational time, these simulations are done in 2D.  To explore the influence of $L$ and $v_{fl}$ on the slowdown of the plasma slabs, we have performed $4$ simulations listed in Table \ref{1} varying these parameters, while keeping $v_{th} \ll v_{fl}$. This parameter space with each of the simulations is presented in Fig.\,\ref{Fig7}.
\begin{table}[ht!]
	\centering
	\caption{Slowdown, visible in the respective measures for final velocity $v_{inst/init}$, expected due to the two-stream/oblique or Weibel instability for various $v_{fl}$ and $L$ with cold slabs $v_{th}/v_{fl}=0.1$.
	*Although the Weibel is not able to saturate, it still causes a significant slowdown.}
	
    	\begin{tabular}{|p{0.7cm} |p{0.77cm}| p{1.208cm}| p{2.1cm}|p{1.4cm}|p{1.4cm}|}
    	
    	\hline
    	Run  &  $v_{fl}/c$ \ & L [$\omega_{p}/c$] & Cause &  $v_{inst}/v_{fl}$ & $v_{init}/v_{fl}$ \\ \hline
		
		$R_{1}$   & 0.1  &   100  &  WI/OBI & 0.5000 & 0.1444 \\ \hline
		
		$R_{2}$   & 0.01  &    100 &   WI/OBI & 0.5320 & 0.1235  \\ \hline
		
		$R_{3}$   & 0.01  &    5 &   WI*/OBI   & 0.651 & 0.442  \\ \hline
		
		$R_{4}$   & 0.01  &  0.02 &   No slowdown & 0.99829 & 0.999976  \\ \hline	
	\end{tabular}\label{1}
\end{table}


For $L/v_{fl} \Gamma_{W} \geq 10$ we expect slowdown of the plasma caused by both the Weibel and oblique instabilities in the red shaded region shown in Fig.  \ref{Fig7}. This is confirmed for runs $R_1$ and $R_2$ which exhibit a significant slowdown. Even when this constraint is not satisfied, and only $L/v_{fl} \Gamma_{TS} \geq 10$, we expect a moderate slowdown caused by the oblique instability shown in Fig.\,\ref{Fig7} in the green shaded region. This is confirmed in $R_3$ where there is moderate slowdown (Some of this slowdown is due to the Weibel instability, even though it does not reach saturation.). However, if neither of these constraints are met, highlighted as the blue region in Fig. \ref{Fig7}, no slowdown is expected. For simulation $R_4$, neither the oblique instability nor the Weibel instability has time to grow and there is no slowdown. For the cases where there is significant slowdown, the velocity is thermalized such that the instantaneous velocity is reduced to the new thermal velocity $v_{inst}=v^{'}_{th}$. For example for Run $R_1$, after the slabs have crossed and $v_{inst}/c=0.05$, the final thermal velocity is $v^{'}_{th}/c = 0.05~(T/m_e c^2 = 0.0025)$.

\begin{table}[ht!]
	\centering
	\caption{Slowdown, visible in the respective measures for final velocity $v_{inst/init}$, expected due to the two-stream/oblique or Weibel instability for various $v_{fl}$ and $L$ with warm slabs $v_{th}/v_{fl}=1.0$. *Although the Weibel is not able to saturate, it still causes a significant slowdown.
	}
    	\begin{tabular}{|p{0.7cm} |p{0.77cm}| p{1.208cm}| p{2.1cm}|p{1.4cm}|p{1.4cm}|}
    	
    	\hline
    	Run  &  $v_{fl}/c$ \ & L [$\omega_{p}/c$] & Cause & $v_{inst}/v_{fl}$ & $v_{init}/v_{fl}$ \\ \hline
    	
    	$R_{5}$   & 0.1  &   100  &  WI & 0.6235 & 0.2980   \\ \hline
    	
    	$R_{6}$   & 0.01  &    100 &   WI & 0.85713 & 0.0564  \\ \hline
    	
    	$R_{7}$   & 0.01  &    5  &   WI*    & 0.90408 & 0.4669   \\ \hline
    	
    	$R_{8}$   & 0.01  &  0.02 &   No slowdown & 0.99988 & 0.99977  \\ \hline	
    \end{tabular}\label{2}
\end{table}

To understand the role of the initial thermal velocity of the slabs, we perform similar simulations with the set-up described in the above section, but with a significant thermal velocity ($v_{th}=v_{fl}$), as shown in Table \ref{2}. 

In a warm plasma ($v_{fl} \le v_{th}$), the two peak structure seen in Fig.\,\ref{Fig2}b, which drives the two-stream/oblique instability, is not present due to significant thermal spread. However, slowdown can occur due to the Weibel instability for $L/v_{fl} \Gamma_{W} \geq 10$.\,Therefore, in this scenario, we do not expect a significant slowdown in both the green and the blue region where the plasma slabs $L (\omega_{p}/c)$ are smaller than 10. This is confirmed in runs $R_7$ and $R_8$ for $v_{inst}$, although for $R_7$ there is still significant slowdown in $v_{init}$ due to the Weibel instability, even though saturation is not reached. The Weibel instability thus significantly thermalizes and mixes the two slabs, but the plasma still exits the collision at the thermal velocity seemingly unaffected. On the other hand, in the red region, for plasma length $L (\omega_{p}/c)$ greater than 100, we always expect slowdown. This is confirmed in runs $R_5$ and $R_6$.

\section{Conclusion}
In conclusion, we have shown that the plasma clouds slow down due to Weibel generated magnetic fields, which deflect the particle trajectories, such that particles acquire transverse momentum, thus leading to an isotropic velocity distribution. This process causes the flow velocity to decrease by $1/\sqrt{3}$ in a time interval $\Delta t_{\alpha B} \omega_p \sim c/(v_{fl}\sqrt{\alpha_B})$ , where $\alpha_B$ is the magnetic equipartition parameter, $v_{fl}$ the initial flow speed, and c the light speed, compared with the plasma instability growth time. We show that if the typical plasma slab length is much longer than $v_{fl} \Delta t_{\alpha B}$, plasma particles are always expected to slow down by a factor of about $1/\sqrt{3}$. However, for a cold plasma it may be sufficient to have $L \sim v_{fl} \Delta t_{\alpha E}$, where the oblique mode is able to saturate, in order to obtain a significant slowdown.

The implication of our findings is not only limited to the collision of two plasma slabs. For instance, it has been recently found out that the head-on collision between two electrostatic shocks leads to the development of the Weibel instability, which causes the shocks to slow down similarly to the plasma slabs \cite{Boella2017}.

Furthermore, our conclusions extend to  other areas of physics where collisionless electromagnetic like interactions are prone to occur \cite{Ackerman09, Feng09, Foot16}. Recently, there is a growing interest in investigating interactions of dark electromagnetism, which behaves like a cold collisionless plasma of self-interacting dark matter particles \cite{Heikinheimo15}. Therefore, plasma instabilities are likely to occur in this setup. Observations of collisions between dark matter populations allow to estimate a constraint on the dark electromagnetic coupling constant. Based our results, we estimate the upper bound on the strength of the dark electromagnetic self-interaction $\alpha_{D}\left(\alpha_{D} \approx 4 \times 10^{-25} \ll 1\right)$. This bound assumes the most basic unbroken U(1) gauge interaction of the dark plasma. The self-interaction of two cold $(v_{fl} \geq v_{th})$ plasma clouds leads to the generation of the Weibel and oblique instabilities, which deflects particle trajectories such that the particles acquire transverse momentum while losing forward momentum.  

Finally, our result is non-relativistic, we believe our study can be extended to the relativistic electron-positron fireball beam interacting with plasma, which is important both in the laboratory \cite{Sari-Nature-2015, NShukla20, Arrowsmith} and in astrophysical contexts \cite{Piran04}.

\begin{acknowledgments}
N. Shukla and K. Schoeffler contributed equally to this work. This work was partially supported by the European Research Council (ERC-2015-InPairs-695088). Simulations were performed on the IST cluster (Lisbon, Portugal) and the supercomputer ARCHER (EPCC, UK) in the framework of the HPC-Europa3 Transnational Access program.
\end{acknowledgments}

%


\begin{thebibliography}{36}%
\makeatletter
\providecommand \@ifxundefined [1]{%
 \@ifx{#1\undefined}
}%
\providecommand \@ifnum [1]{%
 \ifnum #1\expandafter \@firstoftwo
 \else \expandafter \@secondoftwo
 \fi
}%
\providecommand \@ifx [1]{%
 \ifx #1\expandafter \@firstoftwo
 \else \expandafter \@secondoftwo
 \fi
}%
\providecommand \natexlab [1]{#1}%
\providecommand \enquote  [1]{``#1''}%
\providecommand \bibnamefont  [1]{#1}%
\providecommand \bibfnamefont [1]{#1}%
\providecommand \citenamefont [1]{#1}%
\providecommand \href@noop [0]{\@secondoftwo}%
\providecommand \href [0]{\begingroup \@sanitize@url \@href}%
\providecommand \@href[1]{\@@startlink{#1}\@@href}%
\providecommand \@@href[1]{\endgroup#1\@@endlink}%
\providecommand \@sanitize@url [0]{\catcode `\\12\catcode `\$12\catcode
  `\&12\catcode `\#12\catcode `\^12\catcode `\_12\catcode `\%12\relax}%
\providecommand \@@startlink[1]{}%
\providecommand \@@endlink[0]{}%
\providecommand \url  [0]{\begingroup\@sanitize@url \@url }%
\providecommand \@url [1]{\endgroup\@href {#1}{\urlprefix }}%
\providecommand \urlprefix  [0]{URL }%
\providecommand \Eprint [0]{\href }%
\providecommand \doibase [0]{http://dx.doi.org/}%
\providecommand \selectlanguage [0]{\@gobble}%
\providecommand \bibinfo  [0]{\@secondoftwo}%
\providecommand \bibfield  [0]{\@secondoftwo}%
\providecommand \translation [1]{[#1]}%
\providecommand \BibitemOpen [0]{}%
\providecommand \bibitemStop [0]{}%
\providecommand \bibitemNoStop [0]{.\EOS\space}%
\providecommand \EOS [0]{\spacefactor3000\relax}%
\providecommand \BibitemShut  [1]{\csname bibitem#1\endcsname}%
\let\auto@bib@innerbib\@empty
\bibitem [{\citenamefont {{S. Weinberg}}(1972)}]{Weinberg72}%
  \BibitemOpen
  \bibfield  {author} {\bibinfo {author} {\bibnamefont {{S. Weinberg}}},\
  }\href@noop {} {\bibfield  {journal} {\bibinfo  {journal} {Gravitation and
  Cosmology, Wiley. New York}\ } (\bibinfo {year} {1972})}\BibitemShut
  {NoStop}%
\bibitem [{\citenamefont {{ G. W. Gibbons, S. W. Hawking, and S.
  Siklos}}(1983)}]{Gibbons83}%
  \BibitemOpen
  \bibfield  {author} {\bibinfo {author} {\bibnamefont {{ G. W. Gibbons, S. W.
  Hawking, and S. Siklos}}},\ }\href@noop {} {\bibfield  {journal} {\bibinfo
  {journal} {The Very Early Universe (Cambridge University Press)}\ } (\bibinfo
  {year} {1983})}\BibitemShut {NoStop}%
\bibitem [{\citenamefont {{T.Piran}}(199)}]{Piran04}%
  \BibitemOpen
  \bibfield  {author} {\bibinfo {author} {\bibnamefont {{T.Piran}}},\
  }\href@noop {} {\bibfield  {journal} {\bibinfo  {journal} {Phys. Rep}\
  }\textbf {\bibinfo {volume} {314}},\ \bibinfo {pages} {575} (\bibinfo {year}
  {199})}\BibitemShut {NoStop}%
\bibitem [{\citenamefont {{R. Schlickeiser and P. K.
  Shukla}}(03)}]{Schlickeiser03}%
  \BibitemOpen
  \bibfield  {author} {\bibinfo {author} {\bibnamefont {{R. Schlickeiser and P.
  K. Shukla}}},\ }\href@noop {} {\bibfield  {journal} {\bibinfo  {journal}
  {Astrophys. J.}\ }\textbf {\bibinfo {volume} {599}},\ \bibinfo {pages} {L57}
  (\bibinfo {year} {03})}\BibitemShut {NoStop}%
\bibitem [{\citenamefont {{D. A. Uzdensky and S.
  Rightley}}(2014)}]{Uzdensky14}%
  \BibitemOpen
  \bibfield  {author} {\bibinfo {author} {\bibnamefont {{D. A. Uzdensky and S.
  Rightley}}},\ }\href@noop {} {\bibfield  {journal} {\bibinfo  {journal} {Rep.
  Prog. Phys.}\ }\textbf {\bibinfo {volume} {77}},\ \bibinfo {pages} {036902}
  (\bibinfo {year} {2014})}\BibitemShut {NoStop}%
\bibitem [{\citenamefont {{L. W. Widrow}}(2002)}]{Widrow}%
  \BibitemOpen
  \bibfield  {author} {\bibinfo {author} {\bibnamefont {{L. W. Widrow}}},\
  }\href@noop {} {\bibfield  {journal} {\bibinfo  {journal} {Rev. Mod. Phys.}\
  }\textbf {\bibinfo {volume} {74}},\ \bibinfo {pages} {775} (\bibinfo {year}
  {2002})}\BibitemShut {NoStop}%
\bibitem [{\citenamefont {{Kulsrud, R. M, and Zweibel, E.
  G.}}(2008)}]{Kulsrud-2008}%
  \BibitemOpen
  \bibfield  {author} {\bibinfo {author} {\bibnamefont {{Kulsrud, R. M, and
  Zweibel, E. G.}}},\ }\href@noop {} {\bibfield  {journal} {\bibinfo  {journal}
  {Rep. Prog. Phys.}\ }\textbf {\bibinfo {volume} {71}},\ \bibinfo {pages}
  {046901} (\bibinfo {year} {2008})}\BibitemShut {NoStop}%
\bibitem [{\citenamefont {{ M. V. Medvedev and A.
  Leob}}(1999)}]{Medvedev-ApJ-1999}%
  \BibitemOpen
  \bibfield  {author} {\bibinfo {author} {\bibnamefont {{ M. V. Medvedev and A.
  Leob}}},\ }\href@noop {} {\bibfield  {journal} {\bibinfo  {journal}
  {Astrophys. J.}\ }\textbf {\bibinfo {volume} {526}},\ \bibinfo {pages} {697}
  (\bibinfo {year} {1999})}\BibitemShut {NoStop}%
\bibitem [{\citenamefont {Schoeffler}\ \emph {et~al.}(2014)\citenamefont
  {Schoeffler}, \citenamefont {Loureiro}, \citenamefont {Fonseca},\ and\
  \citenamefont {Silva}}]{Schoeffler-2014}%
  \BibitemOpen
  \bibfield  {author} {\bibinfo {author} {\bibfnamefont {K.~M.}\ \bibnamefont
  {Schoeffler}}, \bibinfo {author} {\bibfnamefont {N.~F.}\ \bibnamefont
  {Loureiro}}, \bibinfo {author} {\bibfnamefont {R.~A.}\ \bibnamefont
  {Fonseca}}, \ and\ \bibinfo {author} {\bibfnamefont {L.~O.}\ \bibnamefont
  {Silva}},\ }\href@noop {} {\bibfield  {journal} {\bibinfo  {journal} {Phys.
  Rev. Lett.}\ }\textbf {\bibinfo {volume} {112}},\ \bibinfo {pages} {175001}
  (\bibinfo {year} {2014})}\BibitemShut {NoStop}%
\bibitem [{\citenamefont {{ N Shukla, K Schoeffler, E Boella, J Vieira, R
  Fonseca, and A. Leob}}(2020)}]{Shukla20}%
  \BibitemOpen
  \bibfield  {author} {\bibinfo {author} {\bibnamefont {{ N Shukla, K
  Schoeffler, E Boella, J Vieira, R Fonseca, and L. O. Silva}}},\ }\href@noop {}
  {\bibfield  {journal} {\bibinfo  {journal} {Phys. Rev. Research}\ }\textbf
  {\bibinfo {volume} {2}},\ \bibinfo {pages} {023129} (\bibinfo {year}
  {2020})}\BibitemShut {NoStop}%
\bibitem [{\citenamefont {{ T. Silva, K Schoeffler, J Vieira, M. Hoshino, R
  A. Fonseca, and L. O. Silva}}(2020)}]{Silva20}%
  \BibitemOpen
  \bibfield  {author} {\bibinfo {author} {\bibnamefont {{ T. Silva, K
  Schoeffler, J Vieira, M. Hoshino, R A. Fonseca, and L. O. Silva}}},\ }\href@noop {}
  {\bibfield  {journal} {\bibinfo  {journal} {Phys. Rev. Research}\ }\textbf
  {\bibinfo {volume} {2}},\ \bibinfo {pages} {023080} (\bibinfo {year}
  {2020})}\BibitemShut {NoStop}%
\bibitem [{\citenamefont {{C. D. Arrowsmith, et. al}}(2021)}]{Arrowsmith}%
  \BibitemOpen
  \bibfield  {author} {\bibinfo {author} {\bibnamefont {{C. D. Arrowsmith, et.
  al}}},\ }\href@noop {} {\bibfield  {journal} {\bibinfo  {journal} {Phys. Rev.
  Research}\ }\textbf {\bibinfo {volume} {3}},\ \bibinfo {pages} {023103}
  (\bibinfo {year} {2021})}\BibitemShut {NoStop}%
\bibitem [{\citenamefont {{J. R. Peterson, S. Glenzer, and F. Fiuza
  }}(2021)}]{Peterson21}%
  \BibitemOpen
  \bibfield  {author} {\bibinfo {author} {\bibnamefont {{J. R. Peterson, S.
  Glenzer, and F. Fiuza }}},\ }\href@noop {} {\bibfield  {journal} {\bibinfo
  {journal} {Phys. Rev. Lett}\ }\textbf {\bibinfo {volume} {126}},\ \bibinfo
  {pages} {215101} (\bibinfo {year} {2021})}\BibitemShut {NoStop}%
\bibitem [{\citenamefont {Weibel}(1959)}]{Weibel}%
  \BibitemOpen
  \bibfield  {author} {\bibinfo {author} {\bibfnamefont {E.~S.}\ \bibnamefont
  {Weibel}},\ }\href@noop {} {\bibfield  {journal} {\bibinfo  {journal} {{Phys.
  Rev. Lett.}}\ }\textbf {\bibinfo {volume} {2}},\ \bibinfo {pages} {83}
  (\bibinfo {year} {1959})}\BibitemShut {NoStop}%
\bibitem [{\citenamefont {Fried}(1959)}]{Fried}%
  \BibitemOpen
  \bibfield  {author} {\bibinfo {author} {\bibfnamefont {B.~D.}\ \bibnamefont
  {Fried}},\ }\href@noop {} {\bibfield  {journal} {\bibinfo  {journal} {Phys.
  Fluids}\ }\textbf {\bibinfo {volume} {2}},\ \bibinfo {pages} {337} (\bibinfo
  {year} {1959})}\BibitemShut {NoStop}%
\bibitem [{\citenamefont {Tsilva}(2021)}]{Tsilva}%
  \BibitemOpen
  \bibfield  {author} {\bibinfo {author} {\bibnamefont
  {T. Silva, B. Afeyan and L. O. Silva}},\ }\href@noop {} {\bibfield  {journal} {\bibinfo  {journal} {{Phys.
  Rev. E}}\ }\textbf {\bibinfo {volume} {accepted}},
  (\bibinfo {year} {2021})}\BibitemShut {NoStop}%
\bibitem [{\citenamefont {Stamper}\ \emph {et~al.}(1971)\citenamefont
  {Stamper}, \citenamefont {Papadopoulos}, \citenamefont {Sudan}, \citenamefont
  {Dean}, \citenamefont {McLean},\ and\ \citenamefont {Dawson}}]{Stamper-1971}%
  \BibitemOpen
  \bibfield  {author} {\bibinfo {author} {\bibfnamefont {J.~A.}\ \bibnamefont
  {Stamper}}, \bibinfo {author} {\bibfnamefont {K.}~\bibnamefont
  {Papadopoulos}}, \bibinfo {author} {\bibfnamefont {R.~N.}\ \bibnamefont
  {Sudan}}, \bibinfo {author} {\bibfnamefont {S.~O.}\ \bibnamefont {Dean}},
  \bibinfo {author} {\bibfnamefont {E.~A.}\ \bibnamefont {McLean}}, \ and\
  \bibinfo {author} {\bibfnamefont {J.~M.}\ \bibnamefont {Dawson}},\
  }\href@noop {} {\bibfield  {journal} {\bibinfo  {journal} {Phys. Rev. Lett.}\
  }\textbf {\bibinfo {volume} {26}},\ \bibinfo {pages} {1012} (\bibinfo {year}
  {1971})}\BibitemShut {NoStop}%
\bibitem [{\citenamefont {{P. M. Nilson, et. al}}(2006)}]{Nilson}%
  \BibitemOpen
  \bibfield  {author} {\bibinfo {author} {\bibnamefont {{P. M. Nilson, et.
  al}}},\ }\href@noop {} {\bibfield  {journal} {\bibinfo  {journal} {Phys. Rev.
  Lett.}\ }\textbf {\bibinfo {volume} {97}},\ \bibinfo {pages} {255001}
  (\bibinfo {year} {2006})}\BibitemShut {NoStop}%
\bibitem [{\citenamefont {{G. Sarri, \emph{et. al}}}(2015)}]{Sari-Nature-2015}%
  \BibitemOpen
  \bibfield  {author} {\bibinfo {author} {\bibnamefont {{G. Sarri, \emph{et.
  al}}}},\ }\href@noop {} {\bibfield  {journal} {\bibinfo  {journal} {Nat.
  Commun.}\ }\textbf {\bibinfo {volume} {6}},\ \bibinfo {pages} {6747}
  (\bibinfo {year} {2015})}\BibitemShut {NoStop}%
\bibitem [{\citenamefont {G\"ode}\ \emph {et~al.}(2017)\citenamefont {G\"ode},
  \citenamefont {R\"odel}, \citenamefont {Zeil}, \citenamefont {Mishra},
  \citenamefont {Gauthier}, \citenamefont {Brack}, \citenamefont {Kluge},
  \citenamefont {MacDonald}, \citenamefont {Metzkes}, \citenamefont {Obst},
  \citenamefont {Rehwald}, \citenamefont {Ruyer}, \citenamefont {Schlenvoigt},
  \citenamefont {Schumaker}, \citenamefont {Sommer}, \citenamefont {Cowan},
  \citenamefont {Schramm}, \citenamefont {Glenzer},\ and\ \citenamefont
  {Fiuza}}]{Gode-2017}%
  \BibitemOpen
  \bibfield  {author} {\bibinfo {author} {\bibfnamefont {S.}~\bibnamefont
  {G\"ode}}, \bibinfo {author} {\bibfnamefont {C.}~\bibnamefont {R\"odel}},
  \bibinfo {author} {\bibfnamefont {K.}~\bibnamefont {Zeil}}, \bibinfo {author}
  {\bibfnamefont {R.}~\bibnamefont {Mishra}}, \bibinfo {author} {\bibfnamefont
  {M.}~\bibnamefont {Gauthier}}, \bibinfo {author} {\bibfnamefont {F.-E.}\
  \bibnamefont {Brack}}, \bibinfo {author} {\bibfnamefont {T.}~\bibnamefont
  {Kluge}}, \bibinfo {author} {\bibfnamefont {M.~J.}\ \bibnamefont
  {MacDonald}}, \bibinfo {author} {\bibfnamefont {J.}~\bibnamefont {Metzkes}},
  \bibinfo {author} {\bibfnamefont {L.}~\bibnamefont {Obst}}, \bibinfo {author}
  {\bibfnamefont {M.}~\bibnamefont {Rehwald}}, \bibinfo {author} {\bibfnamefont
  {C.}~\bibnamefont {Ruyer}}, \bibinfo {author} {\bibfnamefont {H.-P.}\
  \bibnamefont {Schlenvoigt}}, \bibinfo {author} {\bibfnamefont
  {W.}~\bibnamefont {Schumaker}}, \bibinfo {author} {\bibfnamefont
  {P.}~\bibnamefont {Sommer}}, \bibinfo {author} {\bibfnamefont {T.~E.}\
  \bibnamefont {Cowan}}, \bibinfo {author} {\bibfnamefont {U.}~\bibnamefont
  {Schramm}}, \bibinfo {author} {\bibfnamefont {S.}~\bibnamefont {Glenzer}}, \
  and\ \bibinfo {author} {\bibfnamefont {F.}~\bibnamefont {Fiuza}},\
  }\href@noop {} {\bibfield  {journal} {\bibinfo  {journal} {Phys. Rev. Lett.}\
  }\textbf {\bibinfo {volume} {118}},\ \bibinfo {pages} {194801} (\bibinfo
  {year} {2017})}\BibitemShut {NoStop}%
\bibitem [{\citenamefont {{L. O. Silva, R. A. Fonseca, J. W. Tonge, W. B. Mori
  and J. M. Dawson}}(2002)}]{Luis-2002}%
  \BibitemOpen
  \bibfield  {author} {\bibinfo {author} {\bibnamefont {{L. O. Silva, R. A.
  Fonseca, J. W. Tonge, W. B. Mori and J. M. Dawson}}},\ }\href@noop {}
  {\bibfield  {journal} {\bibinfo  {journal} {Phys. Plasmas}\ }\textbf
  {\bibinfo {volume} {9}},\ \bibinfo {pages} {2458} (\bibinfo {year}
  {2002})}\BibitemShut {NoStop}%
\bibitem [{\citenamefont {{N. Shukla, A. Stockem, F. Fiuza and L. O.
  Silva}}(2012)}]{Shukla-JPP-2012}%
  \BibitemOpen
  \bibfield  {author} {\bibinfo {author} {\bibnamefont {{N. Shukla, A. Stockem,
  F. Fiuza and L. O. Silva}}},\ }\href@noop {} {\bibfield  {journal} {\bibinfo
  {journal} {J. Plasma Phys.}\ }\textbf {\bibinfo {volume} {78}},\ \bibinfo
  {pages} {181} (\bibinfo {year} {2012})}\BibitemShut {NoStop}%
\bibitem [{\citenamefont {{C. Ruyer, L. Gremillet, G. Bonnaud, and C.
  Riconda}}(2016)}]{Ruyer16}%
  \BibitemOpen
  \bibfield  {author} {\bibinfo {author} {\bibnamefont {{C. Ruyer, L.
  Gremillet, G. Bonnaud, and C. Riconda}}},\ }\href@noop {} {\bibfield
  {journal} {\bibinfo  {journal} {Phys. Rev. Lett}\ }\textbf {\bibinfo {volume}
  {116}},\ \bibinfo {pages} {065001} (\bibinfo {year} {2016})}\BibitemShut
  {NoStop}%
\bibitem [{\citenamefont {{A. Bret, L. Gremillet and M. E.
  Dieckmann}}(2010)}]{Bret10}%
  \BibitemOpen
  \bibfield  {author} {\bibinfo {author} {\bibnamefont {{A. Bret, L. Gremillet
  and M. E. Dieckmann}}},\ }\href@noop {} {\bibfield  {journal} {\bibinfo
  {journal} {Phys. Plasmas}\ }\textbf {\bibinfo {volume} {17}},\ \bibinfo
  {pages} {120501} (\bibinfo {year} {2010})}\BibitemShut {NoStop}%
\bibitem [{\citenamefont {{A. Bret and L. Gremillet}}(2006)}]{Bret2006}%
  \BibitemOpen
  \bibfield  {author} {\bibinfo {author} {\bibnamefont {{A. Bret and L.
  Gremillet}}},\ }\href@noop {} {\bibfield  {journal} {\bibinfo  {journal}
  {Plasma Phys. Control. Fusion}\ }\textbf {\bibinfo {volume} {48}} (\bibinfo
  {year} {2006})}\BibitemShut {NoStop}%
\bibitem [{\citenamefont {Bohm}\ and\ \citenamefont {Gross}(1949)}]{Bohm49}%
  \BibitemOpen
  \bibfield  {author} {\bibinfo {author} {\bibfnamefont {D.}~\bibnamefont
  {Bohm}}\ and\ \bibinfo {author} {\bibfnamefont {E.~P.}\ \bibnamefont
  {Gross}},\ }\href@noop {} {\bibfield  {journal} {\bibinfo  {journal} {Phys.
  Rev.}\ }\textbf {\bibinfo {volume} {75}},\ \bibinfo {pages} {1851} (\bibinfo
  {year} {1949})}\BibitemShut {NoStop}%
\bibitem [{\citenamefont {{Y. B. Fainberg, V. D. Shapiro and V. I.
  Shevchenko}}(1970)}]{Fainberg70}%
  \BibitemOpen
  \bibfield  {author} {\bibinfo {author} {\bibnamefont {{Y. B. Fainberg, V. D.
  Shapiro and V. I. Shevchenko}}},\ }\href@noop {} {\bibfield  {journal}
  {\bibinfo  {journal} {JETPL}\ }\textbf {\bibinfo {volume} {30}},\ \bibinfo
  {pages} {528} (\bibinfo {year} {1970})}\BibitemShut {NoStop}%
\bibitem [{\citenamefont {{N. Shukla, J. Vieira, P. Muggli, G. Sarri, R.
  Fonseca and L.~O. Silva}}(2018)}]{shukla18}%
  \BibitemOpen
  \bibfield  {author} {\bibinfo {author} {\bibnamefont {{N. Shukla, J. Vieira,
  P. Muggli, G. Sarri, R. Fonseca and L.~O. Silva}}},\ }\href@noop {}
  {\bibfield  {journal} {\bibinfo  {journal} {{J. Plasma Phys.}}\ }\textbf
  {\bibinfo {volume} {84}},\ \bibinfo {pages} {905840302} (\bibinfo {year}
  {2018})}\BibitemShut {NoStop}%
\bibitem [{\citenamefont {{R. A. Fonseca, L. O. Silva, J. Tonge, R. G. Hemker,
  J. M. Dawson and W. B. Mori}}(2002)}]{R1}%
  \BibitemOpen
  \bibfield  {author} {\bibinfo {author} {\bibnamefont {{R. A. Fonseca, L. O.
  Silva, J. Tonge, R. G. Hemker, J. M. Dawson and W. B. Mori}}},\ }\href@noop
  {} {\bibfield  {journal} {\bibinfo  {journal} {{IEEE Trans. Plasma Sci.}}\
  }\textbf {\bibinfo {volume} {30}},\ \bibinfo {pages} {28} (\bibinfo {year}
  {2002})}\BibitemShut {NoStop}%
\bibitem [{\citenamefont {{R. A. Fonseca, R. A. Fonseca, S. F. Martins, L. O.
  Silva, J. W. Tonge, and F. S. Tsung and W. B. Mori}}(2008)}]{R2}%
  \BibitemOpen
  \bibfield  {author} {\bibinfo {author} {\bibnamefont {{R. A. Fonseca, R. A.
  Fonseca, S. F. Martins, L. O. Silva, J. W. Tonge, and F. S. Tsung and W. B.
  Mori}}},\ }\href@noop {} {\bibfield  {journal} {\bibinfo  {journal} {{Plasma
  Phys. Control. Fusion}}\ }\textbf {\bibinfo {volume} {50}},\ \bibinfo {pages}
  {124034} (\bibinfo {year} {2008})}\BibitemShut {NoStop}%
\bibitem [{\citenamefont {{L. O. Silva}}(2006)}]{silva_2006}%
  \BibitemOpen
  \bibfield  {author} {\bibinfo {author} {\bibnamefont {{L. O. Silva}}},\
  }\href@noop {} {\bibfield  {journal} {\bibinfo  {journal} {AIP Conf. Ser.}\
  }\textbf {\bibinfo {volume} {856}},\ \bibinfo {pages} {109} (\bibinfo {year}
  {2006})}\BibitemShut {NoStop}%
\bibitem [{\citenamefont {{T. N. Kato and H. Takabe}}(2008)}]{Kato_2008}%
  \BibitemOpen
  \bibfield  {author} {\bibinfo {author} {\bibnamefont {{T. N. Kato and H.
  Takabe}}},\ }\href@noop {} {\bibfield  {journal} {\bibinfo  {journal}
  {Astrophys. J. Lett.}\ }\textbf {\bibinfo {volume} {681}},\ \bibinfo {pages}
  {L93} (\bibinfo {year} {2008})}\BibitemShut {NoStop}%
\bibitem [{\citenamefont {{E. Boella, K. Schoeffler, N. Shukla, G. Lapenta, R.
  Fonseca, L. O. Silva}}(2017)}]{Boella2017}%
  \BibitemOpen
  \bibfield  {author} {\bibinfo {author} {\bibnamefont {{E. Boella, K.
  Schoeffler, N. Shukla, G. Lapenta, R. Fonseca, L. O. Silva}}},\ }\href@noop
  {} {\bibfield  {journal} {\bibinfo  {journal} {{arXiv:1709.05908}}\ }
  (\bibinfo {year} {2017})}\BibitemShut {NoStop}%
\bibitem [{\citenamefont {{L. Ackerman, M. R. Buckley, S. M. Carroll, and M.
  Kamionkowski}}(2009)}]{Ackerman09}%
  \BibitemOpen
  \bibfield  {author} {\bibinfo {author} {\bibnamefont {{L. Ackerman, M. R.
  Buckley, S. M. Carroll, and M. Kamionkowski}}},\ }\href@noop {} {\bibfield
  {journal} {\bibinfo  {journal} {Phys. Rev. D}\ }\textbf {\bibinfo {volume}
  {79}} (\bibinfo {year} {2009})}\BibitemShut {NoStop}%
\bibitem [{\citenamefont {{J. L. Feng, M. Kaplinghat, H. Tu, and H.
  Yu}}(2009)}]{Feng09}%
  \BibitemOpen
  \bibfield  {author} {\bibinfo {author} {\bibnamefont {{J. L. Feng, M.
  Kaplinghat, H. Tu, and H. Yu}}},\ }\href@noop {} {\bibfield  {journal}
  {\bibinfo  {journal} {J Cosmol Astropart P}\ }\textbf {\bibinfo {volume}
  {2009}},\ \bibinfo {pages} {004} (\bibinfo {year} {2009})}\BibitemShut
  {NoStop}%
\bibitem [{\citenamefont {{R. Foot}}(2016)}]{Foot16}%
  \BibitemOpen
  \bibfield  {author} {\bibinfo {author} {\bibnamefont {{R. Foot}}},\
  }\href@noop {} {\bibfield  {journal} {\bibinfo  {journal} {J Cosmol Astropart
  P}\ }\textbf {\bibinfo {volume} {2016}},\ \bibinfo {pages} {011} (\bibinfo
  {year} {2016})}\BibitemShut {NoStop}%
\bibitem [{\citenamefont {{M. Heikinheimo, M. Raidal, C. Spethmann and H.
  Veerm\"ae}}(2015)}]{Heikinheimo15}%
  \BibitemOpen
  \bibfield  {author} {\bibinfo {author} {\bibnamefont {{M. Heikinheimo, M.
  Raidal, C. Spethmann and H. Veerm\"ae}}},\ }\href@noop {} {\bibfield
  {journal} {\bibinfo  {journal} {Phys. Lett. B}\ }\textbf {\bibinfo {volume}
  {749}},\ \bibinfo {pages} {236 } (\bibinfo {year} {2015})}\BibitemShut
  {NoStop}%
\bibitem [{\citenamefont {{N Shukla, SF Martins, P Muggli, J Vieira, LO
  Silva}}(2020)}]{NShukla20}%
  \BibitemOpen
  \bibfield  {author} {\bibinfo {author} {\bibnamefont {{N Shukla, SF Martins,
  P Muggli, J Vieira, LO Silva}}},\ }\href@noop {} {\bibfield  {journal}
  {\bibinfo  {journal} {New J. Phys}\ }\textbf {\bibinfo {volume} {22}},\
  \bibinfo {pages} {013030} (\bibinfo {year} {2020})}\BibitemShut {NoStop}%
\end{thebibliography}
\end{document}